# Computational Chemistry as "Voodoo Quantum Mechanics": models, parameterization, and software

Frédéric Wieber & Alexandre Hocquet
LHSP – Archives Henri Poincaré
UMR 7117 CNRS & Université de Lorraine
91 avenue de la Libération
B.P. 454
54001 NANCY CEDEX
FRANCE

**Abstract**

Computational chemistry grew in a new era of "desktop modeling", which coincided with a growing demand for modeling software, especially from the pharmaceutical industry. Parameterization of models in computational chemistry is an arduous enterprise, and we argue that this activity leads, in this specific context, to tensions among scientists regarding the lack of epistemic transparency of parameterized methods and the software implementing them. To explicit these tensions, we rely on a corpus which is suited for revealing them, namely the Computational Chemistry mailing List (CCL), a professional scientific discussion forum. We relate one flame war from this corpus in order to assess in detail the relationships between modeling methods, parameterization, software and the various forms of their enclosure or disclosure. Our claim is that parameterization issues are a source of epistemic opacity and that this opacity is entangled in methods and software alike. Models and software must be addressed together to understand the epistemological tensions at stake.

**Introduction**

The quotation in the title (Evleth 1993a), taken from a scientific mailing list, the "Computational Chemistry List", is supposed to illustrate the issues of epistemic opacity and "theoretical tinkering" in computational modeling methods in chemistry, their relations to computers and the development and use of software in a scientific milieu.

Computational chemistry is in some respects the heir of quantum chemistry in the age of growing computing power available to computational science, but its lineage also traces back to other scientific and instrumental fields in chemistry (physical organic chemistry, protein chemistry, spectroscopies). As a scientific field, it has already been discussed by numerous authors, from the perspective of the history and philosophy of quantum chemistry (Gavroglu and Simões 2012; Park 2003, 2009) and its more recent transformation into computational quantum chemistry (Lenhard 2014; Fisher 2016a, 2016b), and from the perspective of the development of molecular mechanics force fields in protein chemistry (Wieber 2012).

Computational chemistry emerged in a certain epoch and also a certain context. It began as a relative minor client of the supercomputing resources of the 1960s and 1970s but rapidly grew to become a dominant actor in the scientific computing field (Bolcer and Hermann 1994). In the 1980s and 1990s, when the Personal Computer and the workstation democratized computation in the laboratory, a new era of "desktop modeling" (Johnson and Lenhard 2011) coincided with a growing demand for modeling software, especially from the pharmaceutical industry (Richon 2008). The molecular modeling software became a huge potential market for hardware manufacturers to sell graphics terminals and computing power to the Big Pharma (Hocquet and Wieber 2017). Moreover, in the 1980s, during the Reagan years, the Universities (in the US at least) launched "technology transfer" programs. Spin-off companies were encouraged. Software produced in the University turned out to be viewed as a potential revenue for academia (Berman 2012). Molecular modeling software became a central artifact in computational chemistry. The issue of how to transfer it from the developer to the potential users arose and raised tensions in the community. Computational chemistry software was once "user-oriented". The only people involved with a scientific program developed for implementing a computational chemistry method were the group of scientists that actually coded it and the few colleagues that were scientifically collaborating with the developers. It was now becoming "market-oriented". Software packages were designed, aggregating various methods of various research groups, and they had to be distributed, supported, and maintained. The relations with the pharmaceutical industry, with its culture of secrecy, exacerbated these tensions. This particular context allows us to emphasize that epistemological issues associated with scientific modeling in computational chemistry are not only about methods but also about software.

The relations between computing and scientific activity have been the subject of numerous studies. Themes such as the "computerization of science" (Agar 2006), the mutual shaping of computing and biology (Chow-White and Garcia-Sancho 2011) or the emergence of computerized evidence-based medicine (November 2011) explore their interplay. Yet, the history of software within computational scientific communities has attracted less attention. In his study of a software used for simulating fluid dynamics, Spencer (2015) has nevertheless showed how analyzing the evolution of a program within a research group can lead to better comprehend the evolving workability of computational models and the associated transformations of the practices and contexts of actions of computational scientists.

We advocate, along with Spencer, that scientific software deserves more attention in the epistemological discussions of models in computational science, and we argue that computational chemistry and its particular context is a pivotal case study. As Lenhard has pointed out, one of the dimensions of the transformation of *quantum* to *computational quantum* chemistry is that "the field of computational quantum chemistry became organized in a market-like fashion" (Lenhard 2014, p. 90), the adoption of this new type of organization being driven by the commercialization of software, which has enlarged the community of its users. In his works on the various approximations and idealizations used by computational quantum chemists to study pericyclic reactions, Fischer has shown that computational chemists are in "a position of partial epistemic opacity with respect to the computational processes that produce numerical results" (Fischer 2016a, p. 320), and that what he calls computational diagnostics is a way to unpack computational models "in order to probe the impact of approximations and idealizations on the results" (Fischer 2016b, p. 253). He also adds, in a footnote: "another factor [of epistemic opacity] is ownership and access to proprietary software. […] Perhaps the algorithms will always be partially epistemic opaque for reasons in addition to the automation of the computational processes" (Fischer 2016b, footnote 10 pp. 253-254). We follow Lenhard and Fischer in pointing out the interest of studying software in computational chemistry, and we aim at engaging into discussing the issue of epistemic opacity in the context of software.

Parameterization in computational chemistry is an arduous enterprise, and we argue that this activity leads to tensions among scientists regarding the lack of epistemic transparency of parameterized methods and the software implementing them. To explicit these tensions, we rely on a corpus which is suited for revealing them, namely the Computational Chemistry mailing List (CCL), a professional scientific discussion forum.

The CCL was created in 1991 to gather members of the fledgling community. Topics on the CCL include questions and answers about technical details, personal opinions about hardware or software, commercial software announcements, and also scientific matters (Pisanty and Labanowski 1996). Because it was public, accessible and open to all, the CCL soon became an informal arena in which all the actors involved with computational chemistry could interact: from graduate students to senior researchers, developers and users from academia or the industry, software vendors, people from supercomputing centers, hardware vendors, etc… The CCL was the arena where all the people linked to molecular modeling software one way or another could debate. From a historical point of view, the lively episodes called "flame wars" are the most interesting in what they reveal of controversies. Controversial debates force the actors to leave their formal and polite stance. That is why the CCL is a unique corpus to account for the tensions induced by software in the community (Wieber and Hocquet in preparation).

We have analyzed previously the multiple tensions among the scientists in this community who develop, license, distribute and also use software in intertwined academic and industrial contexts (Hocquet and Wieber, 2017). In the present paper, we emphasize on the epistemological issues at stake, specifically regarding epistemic opacity within the pivotal issue of parameterization. We relate one flame war from our corpus from the standpoint of the epistemological tensions induced by parameterization issues. We first make a brief account of the two different epistemic cultures, quantum chemistry and molecular mechanics, from which computational chemistry descends. These two cultures share unifying parameterization concerns and the scientific methods associated with them are merged in software, conducting to the advent of the common "technical knowledge community" of computational chemistry (Johnson 2009). We then discuss a lively flame war episode of the CCL to emphasize what issues are at stake for a diversity of involved actors, in order to assess in detail the relationships between modeling methods, parameterization, software and the various forms of their enclosure or disclosure. Our claim is that parameterization issues are a source of epistemic opacity and that this opacity is entangled in methods and software alike. Models and software must be addressed together to understand the epistemological tensions at stake.

**Historical perspective on computational chemistry**

Models in computational chemistry have roots in two distinct fields in the history of chemistry. The mathematical modeling of molecules is an idea which came up to chemists long before computers were available. Quantum chemistry is a scientific field that emerged in the late 1920s. Theoretical physicists left quantum chemists with a very practical problem: to imagine theories (and models) to describe the molecules in a way that could be calculable and useful to the chemists (Gavroglu and Simões 2012). They were the heirs of a reductionist world view of micro-physics, and the naming of the most popular quantum chemical approximation ("ab initio") reflects the idea that the scientific consistency of an approximation method is based on the universality of the constructed models which should not have to deal with the tinkering of parameters (Park 2009). In practice, calculating was done with pencil and paper, desk calculators, and subsequently with the use of excess computer time on the first supercomputers during the 1950s (Bolcer and Hermann 1994), with little rewards in terms of how big were the molecules that could be actually calculated (Park 2003).

Parallel to this, in the 1950s and 1960s, and because the computing facilities were attracting the interest of many scientific fields, a different way of producing computational models of molecular structure arose, based on far simpler theoretical grounds, on a classical conception of molecules in chemistry, and developed entirely on the pragmatic idea of tackling the modeling of what is actually computable. Organic chemists and also biophysicists showed interest in a theory based on a Newtonian classical mechanics view of molecules which appeared in the 1950s for the conformational analysis of strained organic molecules (Westheimer 1956). The very writing of molecular formulas as a paper tool (Klein 2003), and its 3D incarnation, the ball-and-stick model (Francoeur 2001), is the core of a chemist's epistemic culture, one that relies on the chemical concepts of atoms and bonds, and not on the microphysical concepts of nuclei and electrons. This simplified Newtonian view was not dissimilar to what chemists knew as a molecule from the 19th century molecular models, and to how spectroscopists theorized the molecular vibrations (Wilson et al. 1955).

The benefit from this simplistic theory was the prospect to compute the structure of molecules ranging from the smallest to the most frequently encountered in organic, biological and pharmaceutical chemistry, by the computational standards of that time (Wieber 2012). It was an ad-hoc modeling, based on tinkering parameters to fit experimental results, yet a tractable one, because of its straightforward computability. This tractability was also to the detriment of the universality of the model: this ad-hoc modeling proved successful for a limited (but meaningful) number of molecular families (like cycloalkanes, peptides, sugars,...) and the necessary parameterization to achieve results was limiting, because consistent parameterization was the lengthiest and the hardest task of the modeling activity. It could only be achieved for limited molecular families. Each scientific team developed and parameterized their own method (a so-called "empirical force field"). Each team relied on different (and often competitive) protocols, based on different (and sometimes incompatible) spectroscopic or thermodynamical results, to actually define their parameters, to iteratively refine them, and to apply their parameterization to specific molecular domains, like for example, small organic chemistry, polymer chemistry, protein chemistry.

Strategies of parameterization were supposed to first define "atom types". Tetrahedral and trigonal carbon for example could not use the same linear, angular and torsional parameters. Allegedly, alkene trigonal carbon and ketone trigonal carbon neither. Defining a set of "atom types" was thus a fragile balance between oversimplification (resulting in poor accuracy of the produced results) and overspecialization (resulting in an exponentially increasing amount of parameters to define, compute, calibrate, fit, benchmark...). Specialization into a molecular family was the only way to mitigate the task. Parameters were also to be validated according to some reference. Results from X-ray diffraction were used to compute crystallographic geometries, that would compare with molecular mechanics geometry optimizations. Another source of geometric parameters could be quantum calculations on benchmark molecules. This lack of consistency regarding parameters validation implies that parameters were not interchangeable from one empirical force field to the other. Yet, one important point is that missing parameters were a common caveat. When parameters were missing, the only workaround to actually make the computation run and not halt, was the use of ad-hoc "guessed" parameters.

The demarcation between quantum chemistry and molecular mechanics became blurred throughout the evolution of their respective fields, especially as the promises of the computer arose. The famous Boulder discourse of 1959 by the renowned quantum chemist Charles Coulson at the Conference on Molecular Quantum Mechanics (Park 2003) is a turning point when Coulson acknowledges a schism between two irreconcilable groups within quantum chemistry. In particular, during the 1960s, the so-called "semiempirical" methods emerged. Their name was itself a pun on the compromise they represented. They were based on quantum calculations and thus formed a part of quantum chemistry, but they shared the concern for feasibility with molecular mechanics (or "empirical" methods, as molecular mechanics were sometimes called). In order to be actually tractable, the quantum methods should be simplified, and above all, parameterized to achieve computability (the most lengthy calculations of the model should be replaced by empirical parameters). Similarly to molecular mechanics, different and sometimes competitive semiempirical methods, based on different parameterizations, appeared in the 1970s.

Semiempirical methods were akin to ab-initio methods because they shared the same theoretical formulations, but parameterization in semiempirical methods shared some concerns with parameterization in molecular mechanics. Different sources of parameters were used, not only regarding geometries, but also thermodynamical or energetic quantities, and the same lack of consistency regarding parameters definitions and validation protocols led to distinct and competitive methods, just like competitive force fields had arisen in molecular mechanics. Similarly to "atom types" in molecular mechanics, the parameterization in semiempirical methods led to the "missing parameter" caveat, and molecules that included some elements (besides the ever-present carbon, hydrogen, nitrogen and oxygen) could not be calculated, or were calculated with "guessed" parameters.

Quantum chemistry as a keyword mentioned in publications, though rapidly growing through the 1980s and the 1990s, is superseded by molecular mechanics by the turn of the 1990s. Molecular Mechanics, by its scientific origin (in physical organic chemistry), by the wider spectrum of studied molecules, became the favored modeling activity on the industrial side at that time and thus the first to give birth to a commercial activity of selling and licensing software in the field (Boyd 1991).

Nevertheless, this plurality of methods (Fischer 2016b), based on different but communicating epistemic cultures, gradually turned to meet each other with the advent of the computer (and the computer software) as a unifying tool. Quantum chemistry and molecular mechanics were far from ignoring each other, and as a matter of fact, the birth of the computational chemistry discipline, as viewed by its founders, encompassed both fields in a wide spectrum of modeling activities ranging from "pure" ab initio quantum modeling to so-called "empirical" molecular mechanics. Pragmatic quantum chemists, using semiempirical methods, were positioned somewhere in the middle, as Richard Counts (then editorialist of the new "Journal of Computer-Aided Molecular Design") defines it (Counts 1987a). In 1974, The COMP "Computers in chemistry" division was created at the American Chemical Society. During the 1980s, the new scientific field named "computational chemistry" emerged in the keywords used in scientific papers, in the conference calls for papers, but also in the academic and industrial research job offers or in the reports from supercomputing centers, "leading to the recognition of a new kind of chemist, different from a theoretical chemist, different from a physical chemist, an organic chemist, a spectroscopist or a biophysicist" (Counts 1987b, p. 95)[1].

**A CCL flame war**

The conversation thread we want to discuss in order to disclose some of the tensions produced by issues of parameters, methods and software within the chemists' community starts on 06/23/1993, with a seemingly innocuous message. Twenty-nine posts from eighteen subscribers will follow for ten days. This thread constitutes the first flame war ever on the CCL. The first message is an announcement. Andy Holder, then Assistant Professor of Computational/Organic Chemistry at the University of Missouri-Kansas City and CEO of a scientific software company named Semichem, Inc.[2], announces the publication of a paper providing results for a new quantum chemistry semiempirical method named "SAM1". In this message, Holder writes down: "This [the paper] is primarily a listing of results for the new method for a vast array of systems. [...] A more complete paper describing the model will be forthcoming" (Holder 1993a). This is the last sentence quoted here which will launch the debate. Graham Hurst (then an employee of the software company Hypercube, Inc[3].) wrote in the second message of the thread: "this [Holder's] post disturbs me..." (Hurst 1993a). Hurst considers that "it will be impossible to independently reproduce these results" because the model leading to the results has not already been published. He adds: "If the method has not yet been published, then the results should not have been accepted for publication since they cannot be verified".

---

[1] Actually, it can be argued that other scientific fields like planned synthesis or database searching may qualify as "computational chemistry", but we limit ourselves to the restrictive meaning we just described, because we believe that the scientists involved in quantum mechanics and molecular mechanics, though belonging to different epistemic cultures, shared common concerns that bind them into a "technical community of knowledge". Especially the issue of software, and the dynamics and tensions linked to it, turns our chosen subfield into a relevant unit to analyze.
[2] Semichem, Inc. was a company founded to commercialize the SAM1 method (and other similar semiempirical methods) embedded in the AMPAC software package.
[3] Hypercube, Inc. is a software company commercializing the multipurpose Hyperchem package. Hyperchem implements many semiempirical and molecular mechanics methods.

Three directions of discussion are opened up in response to Hurst's post. First, the question of how possible it is to verify the validity of the results is discussed as the possibility to reprogram the computational method by oneself. Is the information necessary to reprogram the method available? Mark Thompson, then research scientist at the Pacific Northwest Laboratory and developer of a freely licensed molecular modeling program called Argus, explains the difficulties he had came across while trying to reprogram the MNDO method with only the publications at hand[4]: typos, inconsistencies in quantities units (due to the importation of geometrical or energetical parameters from experimental results in units used by experimentalists like angstroms for distances or kcal/mol for energies), and non-explicit importation of parameters from other semiempirical methods. He concludes his post by asking "Would anyone else out there who has implemented the MNDO-family of methods care to comment on their experiences?" (Thompson 1993a). Other posters comment about the long lasting of parameters errors in the published literature (Rzepa 1993) or the anomaly of sulfur containing molecules claimed in a submitted paper to have been calculated with a program version that officially does not include any sulfur parameter for the AM1 method. While checking how this is possible, Evleth, a researcher at the CNRS French institution in Paris, says he learned that the sulfur parameters had been implicitly imported from MNDO, another method, without any testing at all (Evleth 1993b). Authors of said paper had then to acknowledge retrospectively that both methods were using "mixed AM1-MNDO parameters" for some chemical elements. The issue of difficult attempts at reprogramming methods illustrates the messiness of parameterization. Evleth concludes in another message, the same day, that "many who do semiempirical calculations accept they are voodoo quantum mechanics and one has to go to the right witchdoctor" (Evleth 1993a).

Similarly, on molecular mechanics, Hurst (1993b) then adds his experience of coding various force fields in the Hyperchem package, and his having a hard time obtaining official lists of parameters to refer to. Said parameters are sometimes claimed to be listed in a reference publication (or doctoral thesis) that happens to contain only partial and incomplete specifications.

---

4 MNDO, AM1, and SAM1 are all semiempirical methods of the same "family", in the sense that they are developed in the same research group. Holder is the only poster that belongs to this research group. Thompson attempts to implement MNDO in his "Argus" free software package.

A more general discussion then opens up regarding the issue of the parameters that are used in semiempirical *and* molecular mechanics methods. These parameters, central in the different methods used, are not always made publicly available. They are sometimes hidden away in the source code of the program, which is not always made public. Hurst cites (perhaps apocryphally) Allinger, the father of the MM2 molecular mechanics method: "the only complete specification [of the force field] is the program itself" (Hurst 1993b). It exemplifies the difficulty to get a hand on parameters, the fact that parameters are not always available publicly in publications, and moreover that the parameters available in programs are sometimes enclosed in a non-open code. As one of the participant of the thread writes down: "[...] we should like to know your opinion on the actual trend in commercializing computational packages without source codes. Does this trend encourage the development of science? And also: up to what limit a computational package can be considered as a product of a single research group?" (Adamo 1993).

In the second direction of the discussion, the tension between the world of academic research and the world of scientific software corporations is underlined. In response to Hurst's first message, Holder concedes that it is not always easy to clearly distinguish scientific from entrepreneurial activities. The scientists' implication in scientific software corporations, along with the costs necessary to develop software, clashes with the values (like transparency or reproducibility) the actors associate with science. As Holder puts it: "So, while Dr. Hurst's point is well-taken and fully subscribed to by me both in my capacity as a university researcher and president of Semichem, there is no intention to "hide" anything. I understand the sensitivity of this issue and I am committed to the pursuit of science in an *open atmosphere*. [...] The development of SAM1 is my primary research activity [...], but Semichem is also spending money to develop this method and will be giving it to the scientific community freely. We withhold only our code. [...] It should be noted, however, that *some interests are not scientific, but competitive"* (emphases added) (Holder 1993b).

In the third direction of the discussion, the problem of publication ethics is discussed. The importance of the peer review process in scientific publishing is underlined and some contributors ask if reviewers do a good job when accepting for publication results which have been obtained by a computational method not fully (and openly) described. The question leads more generally to contrast proprietary methods and open scientific literature. As Mark Thompson, writes down: "I feel very strongly that when a new method is developed and implemented that it must pass the peer review process to gain legitimacy in the scientific community, regardless of whether most other scientists care to re-implement that method or not. Proprietary methods are fine, as long as it is openly known that they are proprietary. Results of proprietary methods do not belong in the open scientific literature" (Thompson 1993b). Of course, these three directions of discussion are interrelated.

The sixth message of the thread, written by Douglas Smith (then Assistant Professor of Chemistry at the University of Toledo) is particularly revealing. In this long post, Smith responds point-by-point, using interleaved posting, to Thompson's whole message. The tensions produced by software within the community are interestingly expressed by contrasting how scientists believe they should act with what they actually do. Thompson has written that "good science is that of reproducibility and independent verification" (Thompson 1993b). Smith points out that it is "universally true and accepted" but "rarely followed" (Smith 1993). Smith uses as an example the issue of parameters used in molecular mechanics, which are regularly modified and adjusted for a particular study without being published in the paper relating to that particular study. More generally, the very nature of such a method (and of semiempirical methods) leads to a multiplication of the parameters used without a clear display of which parameters are used when producing such or such results. In practice, computational chemists act in a way that differs from what they say they should do. Thompson has also written: "If the results of a new method are published without sufficiently describing the method to fulfill the above criteria [reproducibility and independent verification], then I personally could not take the results seriously" (Thompson 1993b). Here again, Smith considers that if this position points to "a real problem", it is "utopian and most likely not practical", because of "the proprietary nature of commercial software" (Smith 1993), and because some people use this type of software as a "black box". He then adds: "Besides, who ever said we had to reveal all our secrets and make them readily available and accessible? When software copyrights and patents really provide adequate protection, maybe I will agree with that attitude" (Smith 1993). Finally, if "results of proprietary methods do not belong in the open scientific literature", as Thompson has written, "where do they belong?" Smith replies. According to him, the situation is complicated: "what about the difference between someone in industry who paid for the source code for MacroModel[5] as compared to the academic, such as myself, who only gets binaries? Are my results to be less acceptable because I don't have the absolute method available? Or are the industrial results less acceptable because they can be the results of tweaking the code?" (Smith 1993).

---

5 Macromodel is a software package implementing several molecular mechanics methods, including Allinger's MM2 force field, and adding in-house parameters.

It is worth noting that posters switch easily from semiempirical methods to molecular mechanics and back when they talk about their concerns regarding publications, methods, parameterization and software. These two domains share the same concerns and are bound in similar and sometimes even the same software, even if the theoretical formulations they use are very different.

In Smith's post, the discrepancy between the kind of values (transparency, reproducibility) the actors associate with science and their actual practices associated with computational methods and software is clearly highlighted. The issue of the norms of sound science is in practice difficult to address.

Moreover, computational chemists ask the question of how the difficult and tedious work of programming can be recognized. Can this recognition be obtained by publishing programs or by adequately protecting them ("When software copyrights and patents really provide adequate protection […]" Smith writes)? The complexity of the issue of software copyrights and patents is then stressed in many subsequent posts of the thread.

Some posters try to separate the issue of copyright and patenting, as a software issue, from the issue of transparency of methods and publication, as a science issue. Hurst (1993b) makes the strong claim that "it is important to distinguish "science" from "code"" (his emphasis). The former should "include everything a researcher needs to know to reproduce numbers" and the latter "need not be fully or publicly disclosed". But this dichotomous view is criticized. Balducci (1993), a research associate and systems manager at the university of Texas in Austin, opposes that a lot of work in the code, such as defining molecular geometry (and especially ring structures), belongs to science: "in several cases it would be impossible to even describe (much less to reproduce) the "science" of a method without a clear definition of the structure of the "coded" solution". Mercier, then at Cornell school of medicine, regrets that method coding is often "hardwiring tables of parameters into the code" (Mercier 1993) and make them difficult to comprehend. He proposes the example of modular programming, such as in Mathematica software, as a mean to separate parameters from an enclosed code in order to give the community the possibility to enhance the parameterization of a hard-coded method.

Finally, Fernandes, then an undergraduate student at the University of Waterloo, insists that even an open (in the sense of readable) source code is not sufficient to understand the method and its parameterization. Lack of code annotation and obscure versioning do not help: "what guarantee do we have that G92 (or any of Biosym's or AutoDesk's products) actually do what they are supposed to? Even having the source code just doesn't help... who wants to root through 350 000 lines of someone else's code for any reason?" (Fernandes 1993).

The thread finally dies of attrition after a general sense of uncertainty about what the future holds regarding the relationships between intellectual property notions and the tensions they expressed beforehand.

**Parameters as opacity in methods and software**

The initial problem of the flame war is an epistemological problem associated with a problem of publication ethics: as the details of the model used to produce the published results have not been published, the results cannot be independently reproduced and verified and are then not considered as publishable. Beyond this problem, the tensions revealed by the conversation thread are the symptomatic expression of opacity regarding parameters, methods and software.

Andy Holder's first post speaks about a future paper that will "describe the [SAM1] model" and Hurst's infuriated answer states that "the SAM1 *method* still does not have an *official* reference" (his emphasis on method and official). Two subsequent posters speak of disclosure of "semiempirical parameterization" (Evleth 1993b) and "disclosure of parameters" (St-Amant 1993). As a matter of fact, if the titles in the subject headers of the first posts are about "SAM1 reference", there is a shift in titles towards "full disclosure of methods" after two days of posting, and actors even insist on the issue of "disclosure of programs" in subsequent messages. The question is what exactly has to be disclosed – models, methods, parameters, programs, software – and how to ensure transparency.

The epistemological nature of the models in computational chemistry implies that epistemic transparency is very difficult to reach in practice. The very nature of the models, for example in semiempirical methods like SAM1 or in molecular mechanics approaches like MM2, requires a time consuming work of parameterization. Parameterization poses a problem of reproducibility and transparency. Scientific parameters that are designed to make the model actually work possess their own epistemic problems.

Numerous research groups build molecular mechanics force fields, and the essential work of parameterization in this construction is sometimes developed in a competitive atmosphere. Force field success is measured in terms of parameters efficiency (to produce sound results), consistency (to produce reproducible results) but also workability (to avoid calculation failures because of the absence of parameters that lead to a program halt). In force field construction, some parameters are missing, some generic parameters are designed to replace missing parameters to avoid program halt, and some parameters are lacking a sufficient description in order to know if a parameter is proper or simply fills a hole. The situation of force field multiplicity is even made more complex by the fact that competitive software packages may implement a certain force field differently, especially in the treatment of missing parameters. Furthermore, other research groups may adapt a force field to their particular calculation needs by adding layers of "in-house" parameters.

Semiempirical methods show similar issues due to a similar complex situation. Holder (1993b) says that it is impossible to reproduce the MNDO method "from scratch", because publication is incomplete. Yet, Thompson tried to actually reprogram MNDO "from scratch", without the source code, with only the mere paper in hand. Errors, units discrepancies, and mixed parameters (importation of parameters from a method to another to make the program run) were the difficulties encountered by Thompson and others.

Given this diversity of parameterization and parameters descriptions, methods and their parameters are not consistently disclosed, and this is a source of epistemic opacity. Actors ask whether they could be, for example in publication's supplementary material. In the end, the question actors are asking is whether a fully parameterized method could be properly described and then sufficiently disclosed in a publication.

In practice, this ideal of transparency is hard to reach because parameters are intertwined with the coding of the method. As several actors mention, the parameters are often "hardwired" into the code. In one poster's (quoted) words, "the only full description of the method is the program itself". The entanglement of parameters and code is adding a new layer of opacity in the issue of reproducibility.

A serious issue regarding parameters embedded in code is the fact that the code may not be open, in the sense of not readable. If the code of the program is not open, then executable binaries are what software users get. Parameters, which are in that case hidden in an enclosed program, cannot thus be checked.

So why do developers "withhold the code" in Holder's words (Holder 1993b)? Because software is more than just code, it is also a commodity, and the scientific activity of computational chemistry shares common concerns with the industrial sector of software sales. In a world where software is also a business, issues of intellectual property, or software distribution in general, entwine with the traditional concerns of the scientific world. The difficulty to finance continuous development clashes with scientific ethos concerns. This situation gives rise, for example, to the question of the lack of scientific recognition for software development. Many actors then consider that it is important to gain recognition for software development, and in this regard, to protect and enclose the code is vital, even if methods should be reproducible. Modular programming is cited as a mean to give the community the possibility to disclose parameters while leaving the code enclosed.

In the same vein, Hurst (1993b) claims that "science and code must be separated", but this attempt to separate methods and software is viewed by some posters as vane. He is recalled that the entanglement of both is inextricable. For example, Balducci (1993) advocates that beyond mathematical equations formalizing a model, the mere computational definition of molecular structure to effectively compute the calculations (molecular geometry, and especially ring structure) is intermediate between science and code, and belongs to both.

Is code openness, as a proposal to solve this opacity issue, a solution? The idea that any scientific code should be open is not straightforward. Some posters regard the commercialization of computational chemistry packages as presenting many advantages. First, instead of spending time and energy in building code, the use of already existing code can be viewed as the possibility to dedicate oneself to actual chemistry (publishable) calculations, instead of (unrewarded) programming. Second, financing the computational effort of developing software through software sales is seen in some quarters as the only way to have the possibility that robust, versatile, powerful and efficient software packages even exist. And third, from an industrial user point of view, buying commercial software grants liability from a corporate software vendor in case of something going wrong. Some posters then view software packages as a metaphoric scientific instrument: commercialization is seen as building trust in the scientific community.

There is however no consensus on this matter. Posters cite many examples of the need of an open source code for sound scientific practices, and one of them is precisely the need for parameters verifiability. Tensions proceed from the confrontation of two viewpoints: one concerned with epistemic transparency as scientific ethos, and the other concerned with software as a commodity. Academic publishing, which constitutes the traditional form of academic reward, is central in the actors' ideal concept of the openness of science. Yet, there is a tension between two stances. One is that modeling software, as a scientific tool, should be considered a public tool, and as such, one that belongs to the scientific community, including in its potentiality to be disclosed, enhanced, and maintained. The other is that, software, as a tool developed by a small team, in a commercial context, should be licensed, strict licensing policies helping to keep software stable, which guarantees the production of sound scientific results.

Yet, even if the code is made public, the opacity lies also in the complexity of the programs: verifying programs (and not merely testing results) is very difficult according to Fernandes (1993). It is not easy to assess whether the program behaves as intended by the code developers: lack of code annotations and versioning issues are here again sources of opacity, even in a readable code.

Finally, the opacity also lies in the software package licensing policy that impedes checking of parameters. Smith (1993) evokes the paradoxical issue of industrial or academic users of the same software package. The former get access to the source code and thus have the possibility to tinker force field parameters "in house" and then publish computational results that can be criticized for unsoundness (are the parameters used thoroughly tested?) and lack of traceability (are the modified parameters published?). The latter only get access to binaries and their calculations may be criticized for using a method whose parameters they don't know exactly. Packaging and licensing policies add yet another layer of epistemic opacity.

The flame wars as we have depicted one of them are thus the symptoms of a lack of epistemic transparency regarding methods *and* software. First, the opacity of method's parameters lie in their diversity, and this implies issues regarding publications. Methods are opaque if parameters are unavailable. Second, parameters are intertwined with code, which adds a layer of opacity, be it open or not. Third, programs are themselves opaque because of their complexity and lack of traceability. The final layer of opacity lies in the policies of software as packages. We argue that this epistemic situation, and the tensions implied by parameterization at any level, are constitutive of computational chemistry.

## **Conclusion**

The issue of parameterization as a source of epistemic opacity in computational chemistry is a telling example that models and software must be addressed together in computational science. Interrelations between both imply that transparency and validity of computational methods are complex, and that they are a source of tensions for scientists, in the economic, political and technological context we described.

It is interesting and necessary to discuss the structure, properties and epistemological status of models, as it is common in the philosophy of science. We argue that it is furthermore necessary to understand models in relation with software which embody them, which give them their productivity. In turn, understanding software (in computational sciences) needs to take into account the models they express, that is "the representations of the world" scientists translate in a way the computer can "understand". These representations depend on the communities of scientists involved and the histories of the ways they represent the portion of the world they are interested in (Mahoney 2008). In Mahoney's words, these models, and their translations into software, are "operative representations", which are central to our study.

The complexity of the parameterization is central in the modeling activity. This has to be understood in the context of the calculability problems quantum approaches and molecular mechanics approaches have faced. In the case of molecular mechanics methods, the choice of a certain representation of matter, which is consistent with a classical conception of molecules, also leads to a necessary complex work of parameterization. The choices of sets of parameters, made locally by such or such research group for such or such group of molecules, lead to models whose epistemic transparency is questioned by the actors themselves. What is interesting for our argument is that this lack of transparency of models has repercussion on the status of software: the question of the openness of the source code is for example made more salient knowing the importance of parameterization in modeling. What is also interesting is that there are similar concerns with semiempirical methods, even though the latter are quantum methods. In this regard, molecular mechanics and semiempirical quantum methods share a fundamental epistemic issue, especially since they are bound in similar ways to software, and sometimes even within the same package. The phrasing "Voodoo quantum mechanics", as used by one of the actor of the flame war we have related, ironically highlights the issues of opacity in methods, parameterization and software that literally "possess" computational chemists, who have then to rely on "the right witchdoctor".